\documentstyle[12pt]{article}
\newcommand{\be}{\begin{equation}}
\newcommand{\ee}{\end{equation}}
\newcommand{\ba}{\begin{eqnarray}}
\newcommand{\ea}{\end{eqnarray}}

\date{}

\newcommand{\R}{{\rm R}\!\!\!\!\!{\rm l}}
\newcommand{\grgl}{\:\hbox to -0.2pt{\lower2.5pt\hbox{$\sim$}\hss}
           {\raise3pt\hbox{$>$}}\:}
\newcommand{\klgl}{\:\hbox to -0.2pt{\lower2.5pt\hbox{$\sim$}\hss}
           {\raise3pt\hbox{$<$}}\:}

\begin{document}
\begin{titlepage}
\begin{flushright}
HD--THEP--97--12
\end{flushright}
\vspace{1.5cm}
\begin{center}
{\bf\LARGE Effective Nonlocal Euclidean Gravity}\\
\vspace{1cm}
Christof Wetterich\footnote{e-mail C.Wetterich@thphys.uni-heidelberg.de}\\
\bigskip
Institut  f\"ur Theoretische Physik\\
Universit\"at Heidelberg\\
Philosophenweg 16, D-69120 Heidelberg\\
\vspace{1cm}
\end{center}
\begin{abstract}
A nonlocal form of the effective gravitational action
could cure the unboundedness of euclidean
gravity with Einstein action. On sub-horizon length scales
the modified gravitational
field equations seem compatible with
all present tests of general relativity and
post-Newtonian gravity. They induce
a difference in the effective Newton's constant between regions
of space with vanishing or nonvanishing curvature scalar
(or Ricci tensor).
In cosmology they may lead to a value $\Omega<1$ for the
critical density after inflation. The simplest model
considered here appears to be in  conflict with nucleosynthesis,
but generalizations consistent with all cosmological
observations seem conceivable.

\end{abstract}
\end{titlepage}
\newpage
\setcounter{section}{0}
\section{\protect\Large Stability problem in euclidean gravity}
\setcounter{equation}{0}
It is a longstanding observation that the
Einstein-Hilbert action for gravity is not bounded in
euclidean space. Fundamental problems for the formulation of
quantum gravity are connected to this fact.
This stability problem
in euclidean Einstein gravity can be easily seen if
one expands the
curvature scalar $R$ around flat space $\R\ \ {}^d$
with metric fluctuations
$h_{\mu\nu}=g_{\mu\nu}-\delta_{\mu\nu}$.
\footnote{Our conventions are \cite{Wein}
$R=R_{\mu\nu} g^{\mu\nu}=R^\mu_\mu,\  R_{\mu\nu}=
R_{\mu\lambda\nu}{}^\lambda
,\ R_{\mu\nu\rho}{}^{\lambda}=\partial_\mu\Gamma_{\nu\rho}
{}^\lambda-\partial_\nu\Gamma_{\mu\rho}{}^\lambda+
\Gamma_{\mu\tau}{}^\lambda
\Gamma_{\nu\rho}{}^\tau-\Gamma_{\nu\tau}{}^\lambda
\Gamma_{\mu\rho}{}^\tau,
\ g=\det(g_{\mu\nu})$. As compared to earlier work of the same
author, we use here an opposite sign of the
definition of $R_{\mu\nu\rho}{}^\lambda$. For a
continuation to Minkowski space the signature
will be $(-+++)$, also opposite to earlier work
of the author.} Up to a total derivative
one finds in quadratic order
\ba\label{1}
\sqrt{g}R&=&I_1-I_2\nonumber\\I_1&=&\frac{1}{8}(
\partial^\nu h^{\mu\rho}-\partial^\rho h^{\mu\nu})(\partial_\nu
h_{\mu\rho}-\partial_\rho h_{\mu\nu})\nonumber\\
I_2&=&\frac{1}{4}(\partial_\nu h^{\rho\nu}-\partial^\rho h^\nu_\nu)
(\partial_\mu h^\mu_\rho-\partial_\rho h^\mu_\mu).\ea
One observes that for $d>2$ the euclidean Einstein action
$(M_p=G^{-1/2}\approx10^{19}$ GeV)
\be\label{2}
S_E=\frac{M^2_p}{16\pi}\int d^d x\sqrt{g} R\ee
can become arbitrarily negative for configurations
with arbitrarily large $I_2$.
This becomes obvious by decomposing $h_{\mu\nu}$ into
irreducible representations of the Poincar\'e group
\ba\label{3}
h_{\mu\nu}&=&b_{\mu\nu}+\partial_\mu a_\nu+\partial_\nu a_\mu
+(\partial_\mu\partial_\nu-\frac{1}{d}\delta_{\mu\nu}\partial^2)
\chi+\frac{1}{d}\delta_{\mu\nu}\sigma\nonumber\\
\partial_\mu b^{\mu\nu}&=&0\ , \ \partial_\mu a^\mu=0\ ,\ b_{\mu\nu}
\delta^{\mu\nu}=0\ea
where
\be\label{4}
\sqrt{g} R=\frac{1}{4}\partial^\rho b^{\mu\nu}
\partial_\rho b_{\mu\nu}-
\frac{(d-1)(d-2)}{4d^2}\partial_\mu(\sigma-\partial^2\chi)\partial
^\mu(\sigma-\partial^2\chi).\ee
For $d>2$ the Einstein-action becomes negative semi-definite for
configurations with vanishing $b_{\mu\nu}$. A general gauge fixing term
\be\label{5}
S_{gf}=\frac{M_p^2}{64\pi}\int d^d x
(\eta_1\partial_\nu h^{\rho\nu}-\eta_2
\partial^\rho h^\nu_\nu)(\eta_1\partial_\mu h^\mu_\rho-\eta_2\partial_\rho
h^\mu_\mu)\ee
can move the unboundedness to other irreducible
representations but not cure
the disease.\footnote{For this argument it is important to observe that
$\eta_1=\eta_2$ is not an allowed gauge fixing. Otherwise $\eta_1$
 and $\eta_2$ are arbitrary constants.} The unboundedness of $S_E+S_{gf}$
 can also not be stabilized by higher  powers of $h$ since we can always
consider small but arbitrarily strongly varying $h$.
As a result, no well defined partition function $Z=\int  D g_{\mu\nu}
e^{-(S_E+S_{gf}+S_{gh})}$ can be defined with
the Einstein-action for standard
measures $Dg_{\mu\nu}$ respecting general covariance.\footnote{General
covariance forbids an arbitrary change of sign of the second term on the
r.h.s. of eq. (\ref{4}). This would result for an analytic continuation
from Minkowski space where the conformal mode would be rotated with the opposite
sign as compared to the transversal modes \cite{Haw}.
The precise form of
the ghost action $S_{gh}$ is unimportant for our
problem and we have not
written explicitly the necessary functional
integration over ghost fields.}

By itself, the unboundedness of Einstein's action poses
no severe problem for
the quantization of gravity by a functional integral.
There are many possible extensions of this
action which only influence
the short distance behaviour and
make the action bounded. One of the simplest is the addition of a
term $\int d^d x\sqrt{g} R^2$ with a positive coefficient.
The partition function in presence of sources is
then well defined (in presence of a suitable
ultraviolet regularization) and one may compute the
effective action $\Gamma$ as the generating
functional of the 1PI Green functions. For any
classical action which is bounded
from below, however, the effective action should also be bounded from  
below.\footnote{
This holds after subtraction of a suitable constant.}
The instability problem
reappears now for any effective action for
which only the Einstein term
(\ref{2}) (plus gauge fixing) is relevant for the long distance physics.
The absolute minimum of such an effective action cannot correspond to
flat space. We know from eq. (\ref{4}) that there are necessarily
configurations which have a lower effective action as
compared to $g_{\mu\nu}=\delta_{\mu\nu}$. The absolute
minimum of the (bounded) effective action should therefore
correspond to a nonvanishing curvature tensor.
Despite the well known classical stability for small
fluctuations around Minkowski space
(positive energy theorem) there remains
the open question why the observed
ground state of gravity (flat space) is not given by
the minimum of the euclidean effective action. Interpreting as usual
$\exp(-\Gamma)$ as  a probability distribution
the flat space configuration
$g_{\mu\nu}=\delta_{\mu\nu}$ would be
exponentially suppressed as compared to the absolute minimum of
$\Gamma$. With a
normalization $\Gamma[g_{\mu\nu}=\delta_{\mu\nu}]=0$ the latter would
necessarily occur for negative $\Gamma$ and have non-zero curvature.

In this letter
we adopt the attitude that the  effective euclidean
action for gravity should
have its absolute minimum for flat space (in absence of matter).
This means that the long-distance behaviour of the effective
action needs an extension beyond Einstein gravity. We will
investigate a nonlocal effective action for gravity and
discuss possible observational consequences. In view of the
fundamental character of the euclidean stability problem it seems
justified to ask which extensions of Einstein gravity are
allowed phenomenologically without presenting a quantum field
theoretical justification of the assumed nonlocalities.

\section{\protect\Large Nonlocal gravity}

How could an effective action for
gravity at long distances look like?
We will require here general covariance and
positivity for the euclidean action,
with $\Gamma=0$ for flat space.\footnote{We do not
include in our discussion the generalized
gauge fixing term which is also present in the effective action and
constrained by Ward identitites.}
By long distances we mean momentum scales much smaller than $M_p$.  It is
rather obvious that no local effective action can fulfil both conditions
simultaneously. Adding a cosmological constant
term only worsens the stability
problem. The Einstein-Hilbert-term and the
cosmological  constant are the only
invariants involving up to two  derivatives.
Higher derivative terms $\sim R^2,
\ R_{\mu\nu}R^{\mu\nu}$ or $R_{\mu\nu\lambda\sigma}
R^{\mu\nu\lambda\sigma}$ are
ineffective at long distances. They do not remedy
the stability problem either.
This generalizes to any finite number of  derivatives.

One way out of the dilemma seems a nonlocal
form of the effective action.
Since there are always massless graviton
fluctuations around  flat space there
is a priori no argument why the effective action
must be local. Without a consistent nonperturbative
formulation of euclidean gravity with an Einstein action
we can also not rely on the results of a one-loop calculation
of the effective action which is based on a classical
Einstein action \cite{LOOP}. In this letter we explore
the possibility of a nonlocal effective action which
extends Einstein's gravity in a simple way. We are interested
in the general consequences of such an extension, without
having at the moment indications that quantum gravity
really produces the nonlocal features discussed here. Our
model is chosen for simplicity, and we have no arguments which
favour it as compared to more elaborate nonlocal extensions
which we will briefly discuss at the end.
For definiteness we consider
\ba\label{6}
\Gamma&=&\frac{M^2_p}{16\pi}\int d^d x\left\{\sqrt{g} R+\frac{1}{2}
\tau^2{\cal L}_{nl}\right\}\nonumber\\[0.2cm]
{\cal L}_{nl}&=&\sqrt{g}R{\cal D}
^{-1}R=-\sqrt{g} R(\varepsilon R+D^2)^{-1} R\ea
with $D^2$ the covariant Laplacian\footnote{Here $D_\mu S_{\nu\rho
\ldots}=S_{\nu\rho\ldots;\mu}$ denotes a covariant derivative.}
$D^2 R=R;{}^\mu{}_\mu$. We remind here that
the inverse of differential operators has to be handled
with some care and ${\cal D}=-\varepsilon R-D^2$ should only
be considered as an appropriate limit for a properly regularized
operator $\cal D$.
For a given metric $g_{\mu\nu}$ the  meaning of
${\cal D}^{-1}R$ is specified in a basis of eigenvectors of
${\cal D}$
\be\label{6a}
R(x)=\sum_na_n R_n(x)\ ,\ {\cal D}R_n(x)=\lambda_n R_n(x)\ ,\
{\cal D}^{-1} R(x)=\sum_n a_n\lambda_n^{-1} R_n(x).\ee
A regularized operator should have the property
${\cal D}R=0\Rightarrow R=0$ such that all $\lambda_n$
are different from zero and $R=0\Rightarrow {\cal D}^{-1}R=0$.
Translating ${\cal D}=-\varepsilon R-D^2$
to a Fourier basis one finds for $p^2\not=
0$ and $h_{\mu\nu}
\to 0$ that in linear order in $h_{\mu\nu}$ the term
$\sim\varepsilon R$ is
ineffective
\be\label{6b}
(-(D^2+\varepsilon R)^{-1} R)(p)=(-(D^2)^{-1} R)(p)=-
\frac{p^\mu p^\nu}{p^2}(h_{\mu\nu}(p)-h^\rho_\rho(p)
\delta_{\mu\nu})=
\frac{R(p)}{p^2}.\ee
The term $\sim\varepsilon$ plays, however, a role in the limit
where $D^2 R\to 0$ while
$R\not= 0$, as for the example of a sphere. In
this limit $(\varepsilon R+D^2)
^{-1}R$ approaches the constant $1/\varepsilon$ and
this remains so if the radius of the sphere goes to infinity.
Obviously, the two limits are
distinguished by the relative size of $D^2R$ as
compared to $R^2$. We note that
the operator ${\cal D}^{-1}$ still
needs to be regularized for metrics
for which
$D^2R=-\varepsilon R^2$. For our purposes the regularization
\be\label{6c}
{\cal D}=-\varepsilon R-D^2+\frac{\gamma}{M_p^2}
R_{\mu\nu\sigma\lambda}R^{\mu\nu\sigma\lambda}\ee
will be sufficient\footnote{More general regularizations
could easily be written down and ${\cal D}$ could be turned into
a strictly positive operator. We note
that ${\cal L}_{nl}$ is uniquely
defined also for a flat metric or constant $h_{\mu\nu}$
where it vanishes.}, and we omit the regulator term $\sim\gamma$ unless
explicitly needed.

Assuming $\tau^2>0$ the new nonlocal piece is positive
semi-definite for small fluctuations. In terms of
irreducible representations one finds now in quadratic
order in $h_{\mu\nu}$
\ba\label{7}
\sqrt{g}R&+&\frac{1}{2}\tau^2{\cal L}_{nl}=\frac{1}
{4}\partial^\rho b^{\mu\nu}
\partial_\rho b_{\mu\nu}\nonumber\\
&&+\left\{\frac{1}{2}\tau^2(1-\frac{1}{d})^2-
\frac{(d-1)(d-2)}{4d^2}\right\}
\partial^\mu(\sigma-\partial^2\chi)
\partial_\mu(\sigma-\partial^2\chi).\ea
Provided
\be\label{8}
\tau^2\geq\frac{d-2}{2(d-1)}\ee
this expression has indeed its absolute
minimum for flat space! It is therefore
a candidate for an effective action
consistent with our conditions. We note
that  expressed in terms of irreducible
representations  ${\cal L}_{nl}$ is
actually a local expression which takes
the form of an additional positive
kinetic term for the scalar $\sigma-\partial^2
\chi$. In consequence, this
scalar ceases to be a ``tachyon'' (with negative
sign of the euclidean
kinetic term). In other words, $\Gamma$ contains
in quadratic order only standard
kinetic terms for a traceless transversal symmetric
second rank tensor
$b_{\mu\nu}$ and a scalar $\sigma-\partial^2\chi$.
In this language it is
particularly easy to see that the limit
of flat space $(b_{\mu\nu}\to 0,\
\sigma\to 0,\ \chi\to 0)$ is well behaved despite the
operator $(-\varepsilon
R-D^2)^{-1}$ in (\ref{6}).

The linearized gravitational equations (Newtonian limit) can be obtained
if we add to the effective action
a source term $\Gamma_T=\int d^dx{\cal L}_T$ for the conserved
linearized energy momentum tensor
\ba\label{11a}
&&T_{\mu\nu}=V_{\mu\nu}+(\partial_\mu\partial_\nu-\partial^2
\delta_{\mu\nu})W\ ,\  T^\mu_\mu=-(d-1)\partial^2W\nonumber\\
&&\partial^\nu V_{\mu\nu}=0\ ,\ \delta^{\mu\nu}V_{\mu\nu}=0\ea
With $T^{\mu\nu}=-2g^{-1/2}\delta\Gamma_T/\delta g_{\mu\nu}$
one has in the lowest order (upon partial integration)
\be\label{11b}
{\cal L}_T=-\frac{1}{2}h_{\mu\nu}T^{\mu\nu}=-\frac{1}{2}\left
\{b_{\mu\nu}V^{\mu\nu}-(1-\frac{1}{d})(\sigma-\partial^2\chi)
\partial^2W\right\}\ee
and therefore the field equations read for $d=4$
\ba\label{11c}
&&\partial^2 b^{\mu\nu}=-\frac{16\pi}{M_p^2}V^{\mu\nu}\ ,\
\partial^2\tilde\sigma=\frac{16\pi}{M_p^2}\frac{2}{3\tau^2-1}
\partial^2W\nonumber\\
&&\tilde \sigma=\sigma-\partial^2\chi\ea
Only the field equation for $\tilde\sigma$ in presence of
matter $(T^\mu_\mu\not=0)$ is influenced by the nonlocal
term $\sim\tau$. Continuation to Minkowski space changes
the definition of the fields (replacing $\delta_{\mu\nu}$
by $\eta_{\mu\nu}=diag(-1,+1,+1,+1)$
in eqs. (3) (\ref{11a})), but does not affect
the field equations (\ref{11c}). The field equation (\ref{11c}) can
be cast in the form of the usual linearized Einstein equation if we
absorb the contribution from ${\cal L}_{nl}$ into a new
gravitational contribution to the ``total'' energy momentum
$\hat T^{\mu\nu}=T^{\mu\nu}+T^{\mu\nu}_{nl}$,
where $V^{\mu\nu}_{nl}=0$
and $W_{nl}$ can be inferred from $\partial^2\tilde\sigma
=-\frac{32\pi}{M_p^2}(\partial^2 W+\partial^2 W_{nl})$ as
\be\label{11d}
\partial^2 W_{nl}=-\frac{3\tau^2}{3\tau^2-1}\partial^2W\ee
In particular, one finds for a static configuration
\be\label{15a}
\hat T_{00}=T_{00}+\Delta W_{nl}
=T_{00}+\frac{\tau^2}{3\tau^2-1}T^\mu_\mu,\ee
and, for $\sum_iT_{ii}=0$,
\be\label{11e}
\hat T_{00}=\frac{2\tau^2-1}{3\tau^2-1}T_{00}\ee
Up to a rescaling of the effective Newton's constant
\be\label{16a}
G_{N,eff}=\frac{2\tau^2-1}{3\tau^2-1}\frac{1}{M_p^2}=\frac{1}
{(M_{p,eff}^{(0)})^2}\ee
this is standard Newtonian gravity provided $G_{N,eff}$ is
positive. In our context we will impose the condition
\be\label{16b}
\tau^2>\frac{1}{2}\ee
Observation of nonlocal gravity at the linearized level requires
a measurement of the additional coupling to $T^i_i$ which appears
in $\hat T_{00}$ (\ref{15a}) for nonvanishing pressure. At present
we are not aware of such a measurement.

Gravitational radiation in empty space or the propagation
of gravitons is not affected by nonlocal gravity. Gravitons
are described by $b_{\mu\nu}$, and we observe that the source
term for gravitons obtains no nonlocal correction in linearized
order (eq. (\ref{14})). For $\tau\not=0$
this implies that the value of $G_N$ as measured by the emission
of gravitational waves should differ from the static value of
the effective Newton's constant (\ref{16a}). This is, however,
not tested today. Beyond linearized gravity we have to solve
the gravitational field equations for nonlocal gravity.

\section{\protect\Large Gravitational field equations}

The gravitational field equations obtain from
the functional derivative of the effective
action $\Gamma$ with respect to the metric. For
their derivation it is
useful to write $\Gamma$ in an equivalent
local form by introducing a scalar field  $\varphi$
\ba\label{8a}
\varphi&=&-\tau{\cal D}^{-1}R\nonumber\\
\Gamma&=&\frac{M_p^2}{16\pi}\int d^dx\sqrt g\left\{
-\frac{1}{2}\varphi{\cal D}
\varphi+(1-\tau\varphi)R\right\}\ea
Inserting the field equation for $\varphi$,
${\cal D}\varphi=-\tau R$, into
$\Gamma$ one recovers the original
effective action (\ref{6}). Solutions of
the field equations from (\ref{8a}) or (\ref{6}) are therefore
equivalent.\footnote{Note that the general local solution of
${\cal D}\varphi=-\tau R$ has free integration
constants which must be fixed by
appropriate boundary conditions according to (\ref{6a}).
The integration constants are related to a function $\psi$
obeying ${\cal D}\psi=0$ that can be added to the local
solution for $\varphi$. In particular, the function
$\psi$ must be chosen such that
$\varphi$ vanishes for $\tau=0$ or for a uniformly
vanishing curvature scalar. Also
$\varphi$ should have no singularity if $R$ is
nonsingular. More
formally, $\varphi$ must belong to a space
of functions for which
${\cal D}\varphi=0\Rightarrow\varphi=0$.}
For ${\cal D}=-\epsilon R-D^2$ the field
equation for $\varphi$ and
the effective action are given by
\ba\label{9}
&&D^2\varphi=(\tau-\varepsilon\varphi)R\nonumber\\
&&\Gamma=\frac{M^2_p}{16\pi}\int d^d x\sqrt{g}\left\{-\frac{1}{2}
D^\mu\varphi D_\mu\varphi+(1-\tau\varphi+\frac{1}{2}
\varepsilon\varphi^2)R
\right\}.\ea
We emphasize that the form (\ref{9}) should
only be considered as an auxiliary action
for on-shell $\varphi$ obeying (\ref{9}).
For off-shell fluctuations of $\varphi$
the wrong sign kinetic term would produce
new instabilities. On the other hand,
for on-shell $\varphi$ one has, of course,
a contribution which is positive
definite for positive eigenvalues of $(-\varepsilon R-D^2)$
\be\label{11}
\frac{1}{2}\tau^2{\cal L}_{nl}=\frac{1}{2}
\sqrt{g}(D^\mu\varphi D_\mu\varphi
-\varepsilon R\varphi^2).\ee
The gravitational field equations can now be found
by variation of (\ref{9}) with respect
to the metric
\ba\label{15}
R_{\mu\nu}-\frac{1}{2} R g_{\mu\nu}&=&
-\frac{1}{1-\tau\varphi+\frac{1}{2}\varepsilon\varphi^2}
\left\{\left(-\frac{1}{2}+\varepsilon\right)
D_\mu\varphi D_\nu\varphi
+\left(\frac{1}{4}-\varepsilon\right)
D^\rho\varphi D_\rho\varphi\  g_{\mu\nu}
\right.\nonumber\\
&&\left.+(\varepsilon\varphi-\tau)(D_\nu
D_\mu \varphi-D^2\varphi\  g_{\mu\nu})
+\frac{8\pi}{M^2_p} T_{\mu\nu}\right\}.\ea
Here we have added the energy momentum tensor $T_{\mu\nu}$
for matter or radiation.
With (\ref{9}) this implies for the curvature scalar for $d=4$
\ba\label{16}
R&=&\frac{8\pi}{M^2_p}\left[1-3\tau^2-(1-6\varepsilon)\left(\tau-
\frac{1}{2}\varepsilon\varphi\right)\varphi\right]^{-1}\nonumber\\
&&\left( T^\mu_\mu+\frac{M^2_p}{16\pi}
(1-6\varepsilon) D_\mu\varphi D^\mu\varphi\right)\ea
and therefore
\be\label{17}
D^2\varphi=\frac{8\pi}{M^2_p}\quad\frac{\varepsilon\varphi
-\tau}
{3\tau^2-1+(1-6\varepsilon)(\tau-\frac{1}{2}
\varepsilon\varphi)\varphi}
\left( T^\mu_\mu+\frac{M^2_p}{16\pi}(1-6\varepsilon)
D^\mu\varphi D_\mu\varphi
\right).\ee
It is easy to check that eq. (\ref{15})
with $D^2\varphi=(\tau-\varepsilon\varphi)
R$ implies $T^{\mu\nu}{}_{;\nu}=0$.
The field equations are the same for Euclidean
or Minkowski signature of the metric.

In this language it is obvious that the
nonlocal term ${\cal L}_{nl}$
remains compatible with all solutions of
Einstein's field equation in empty space.
This means that an arbitrary solution of
the Einstein-equation with vanishing
energy momentum tensor, $R_{\mu\nu}-\frac{1}{2}
R g_{\mu\nu}=0$, remains a
solution of nonlocal gravity. In fact, for $R=0$
the field equation (\ref{15})
has always the trivial solution $\varphi=0$.
The additional terms $\sim\tau$
in the gravitational field equation (\ref{15}) vanish
for $\varphi=0$ and do therefore
not affect the solutions of the Einstein-equation.
In particular, there are no
corrections to post Newtonian gravity in
contrast to Brans-Dicke-theory or general
scalar-tensor theories.
For example, far away from an extended massive object one
has the standard Schwarzschild metric.

There is,
however, one notable difference from Einstein gravity,
namely that the total energy of the object, which appears
in the integration constant $M$ of the Schwarzschild
solution, gets a modified contribution from the
gravitational energy. It is given by
\be\label{17a}
M=\int_Vd^3x(T_{00}+t_{oo})\ee
with $t_{00}$ defined implicitly by linearizing eq. (\ref{15})
in an asymptotically Minkowskian coordinate system \cite{Wein}
\be\label{17b}
R^{(1)}_{00}-\frac{1}{2}\eta_{00}R^{(1)}=-\frac{8\pi}{M_p^2}(
T_{00}+t_{00}).\ee
The nonlocal contribution to $t_{00}$ beyond Einstein gravity
is found from eq. (\ref{15}) for static $\varphi$
\ba\label{24a}
&&t_{nl,00}=\frac{\varphi(\tau-\frac{1}{2}\epsilon\varphi)}
{1-\tau\varphi+\frac{1}{2}\epsilon\varphi^2}T_{00}-\nonumber\\
&&-\frac{(\tau-\epsilon\varphi)^2T^\mu_\mu-\frac{M^2_p}{32\pi}
[\tau^2-1+4\epsilon+(1-6\epsilon)(\tau-\frac{1}{2}\epsilon\varphi)
\varphi]\partial^i\varphi\partial_i\varphi}
{(1-\tau\varphi+\frac{1}{2}\epsilon\varphi^2)[3\tau^2-1+(1-6\epsilon)
(\tau-\frac{1}{2}\epsilon\varphi)\varphi]}g_{00}\ea
One recovers the Newtonian limit (\ref{11e}) for $\varphi\to0$
and $g_{00}\to-1$. We observe that eq. (\ref{17}) also admits the
constant solution $\varphi=\tau/\epsilon$ for which one finds
$t_{nl,00}=-\frac{\tau^2}{\tau^2-2\epsilon}T_{00}$. In this
regime the nonlocal gravitational energy has a tendency to
cancel the normal part $T_{00}$ for $\epsilon<0$ and to enhance
it for $\epsilon>\frac{1}{2}\tau^2$.

We conclude that the difference between nonlocal gravity and Einstein
gravity manifests itself only in different effective couplings
of the metric to the energy-momentum tensor -- in empty space
both are identical. For a computation of the modified coupling we may
distinguish two regimes, for $\epsilon R$ or $D^2$
dominating the operator $\cal D$. A rough criterion for the
domination of the $\epsilon R$ term (i.e. $\varphi\approx \tau/\epsilon)$
is $\rho/M^2_p\gg L^{-2}$ with $\rho$ the mass density and $L^{-1}$
a typical gradient  of the mass distribution. For $m\approx
\rho L^3$ the total mass of the system this translates into a
bound for the density $\rho/M^4_p\gg M_p^2/m^2$. For $m$ of
the order of a solar mass the critical density
$\rho_c\approx(1\ {\rm GeV})^4$ is roughly of the order
of nuclear density. Without going into more details, it
seems plausible that ordinary stars can well be described
by Newtonian gravity with the effective coupling (\ref{16a}).
Modifications of the gravitational coupling to neutron stars are
possible and we expect very sizable changes for the interior of
black holes. For $\epsilon<\frac{1}{2}\tau^2$ the effective
gravitational coupling  to mass in the interior of a black hole
is weaker than in Einstein gravity and the attractive force
therefore reduced. It would be interesting to investigate possible
effects on the singularity in the center of a black hole.

\section{\protect\Large Cosmology}

Since
the present version of nonlocal gravity is compatible with those
tests of general relativity which are based on solutions
of Einstein equations in empty space, it can only be detected
by anomalies in the gravitational coupling to matter. A promising
test is cosmology.
On cosmological scales space is not empty and
the curvature scalar does not always vanish.
For $R\not=0$ there is a source for $\varphi$
(eq. (\ref{9})) and nonlocal gravity
leads to modifications of the cosmological equations.
For a general homogenous
and isotropic metric with scale factor $a(t)$ one
has (for Minkowski signature (-+++), zero spatial curvature
$k=0, d=4$ and $H=\dot a/a$)
\ba\label{12}
R_{00}&=&3\ddot a/a=3(H^2+\dot H)\nonumber\\
R_{ij}&=&-(\ddot a/a+2\dot a^2/a^2) g_{ij}=-(3H^2+\dot H)
g_{ij}\nonumber\\
R&=&-6(\dot a^2/a^2+\ddot a/a)=-(12 H^2+6\dot H)\nonumber\\
\varphi_{;00}&=&\ddot\varphi\ ,\ \varphi_{;ij}=
-H\dot\varphi g_{ij}.\ea
This yields the field equations
\ba\label{13}
D^2\varphi&=&-(\ddot\varphi+3H\dot\varphi)=
-a^{-3}\frac{d}{dt}(\dot\varphi a^3)
\nonumber\\
&=&-(\varepsilon\varphi-\tau)R=6(\varepsilon
\varphi-\tau)(2H^2+\dot H)\ea
\be\label{14}
H^2=\frac{1}{(1-\tau\varphi+\frac{1}{2}\varepsilon\varphi^2)}
\left\{\frac{8\pi}{3M^2_p}\rho-\frac{1}{12}\dot\varphi^2+
(\tau-\varepsilon\varphi)H\dot\varphi\right\}\ee
\vspace{.3cm}
\be\label{21}
\dot\rho+nH\rho=0\ee
with $n=3$ or 4 for a matter or radiation
dominated universe. Equation (\ref{17})
now reads
\be\label{22}
\ddot\varphi+3H\dot\varphi=\frac{1}{2}\quad
\frac{(\varepsilon\varphi-\tau)
(T+(1-6\varepsilon)\dot\varphi^2)}{3\tau^2-1+
(1-6\varepsilon)(\tau-
\frac{1}{2}\varepsilon\varphi)\varphi}\ee
with
\be\label{23}
T=-\frac{16\pi}{M^2_p} T^\mu_\mu=\frac{16\pi}{M^2_p}(4-n)\rho,\ee
and we remind that $\tau^2>\frac{1}{2} (\ref{16b}).$

For the radiation dominated universe with $n=4$ and $T=0$
we recover the standard
Friedman solution
\be\label{18}
H=\frac{1}{2} t^{-1}\ ,\ \rho=\rho_0 t^{-2}\ ,
\ \varphi=0\ ,\quad
\rho_0=\frac{3M^2_p}{32\pi}.\ee
This is easily understood since for this solution
the curvature scalar vanishes.
There are  therefore no corrections from the
nonlocal term in the effective
action. As an important consequence all cosmological
predictions from the
radiation dominated epoch
are only modified by the difference between $M_p$ and $M^{(0)}_{p
\ \ eff}$. For nucleosynthesis this implies that
for a given temperature $H^2$
is larger than its value in Einstein gravity by a factor
$(3\tau^2-1)/(2\tau^2-1)>\frac{3}{2}$. Such
a big change in the gravitational clock seems not compatible
with the successful cosmological explanation of nucleosynthesis
and the solution (\ref{18}) with $\varphi=0$ is excluded!
We remark, however, that this
is not the only solution of the system of differential equations
(\ref{14})-(\ref{16}). For example, any constant $\varphi$ would
correspond to a solution, but with a modified critical density $\rho_0$
or, equivalently, a modified effective $M_p$.
(These are actually stable asymptotic solutions for initial
conditions with $\dot\varphi\not=0$.) Nevertheless, only
$\varphi=0$
is compatible with the regularization of ${\cal D}$ if the curvature
scalar vanishes identically. Below we will come back to possible
effects of a nonvanishing $R$ during the radiation-dominated period.

The situation
changes for the epoch before the
universe was radiation dominated (inflation),
and, most striking, for the
matter dominated period. The curvature
scalar does not vanish in these cases and
$\varphi=0$ remains not a solution anymore.
We observe that the
system (\ref{13}) - (\ref{15}) exhibits
a solution with static $\varphi$, namely
\be\label{25}
\varphi=\frac{\tau}{\epsilon}\ee
This simply results in a multiplication of
the effective $M^2_p$ by a factor $1-\frac{\tau^2}{2\epsilon}$
\be\label{25a}
(M^{(c)}_{p,eff})^2=M_p^2(1-\frac{\tau^2}{2\epsilon})\ee
For $\epsilon>\frac{1}{2}\tau^2$ or $\epsilon<0$ this
implies for the matter dominated epoch a solution
\be\label{26}
H=\frac{2}{3}t^{-1},\quad\rho=\frac{16}{9}
(1-\frac{\tau^2}{2\epsilon})\rho_0t^{-2}\ee
As in standard matter dominated cosmology
one has $D^2R=2t^{-2}R$.
The solution $\varphi/\tau=(\varepsilon R+D^2)
^{-1}R=1/\varepsilon$
seems somewhat surprising since it is the same
as for a covariantly constant curvature scalar. It is, however,
the general result whenever the operator
$(R+\frac{1}{\varepsilon} D^2)^{-1}=(R+x)^{-1}$
can be expanded in the differential operator $x\ (x\cdot 1=0)$,
i.e. $(R+x)^{-1}R=(1-\frac{1}{R}x+\frac{1}{R}
x\frac{1}{R}x-...)1=1$. Of course, this holds only
approximately in the limit where
the regulator term $\sim \gamma$ in (\ref{6c})
is neglected.\footnote{For $\gamma>0$ one has
with ${\cal D}=y-D^2$ the expansion $(y-D^2)^{-1}
R=-\frac{1}{\varepsilon}\left\{1
-\frac{\gamma}{M_p^2}(1+y^{-1}D^2+(y^{-1}D^2)^2
+(y^{-1}D^2)^3+...)y^{-1}R_{\mu\nu\sigma\lambda}
R^{\mu\nu\sigma\lambda}\right\}$.
For $H=\eta t^{-1}$ and $|\varepsilon R|\gg(\gamma/M_p^2)
R_{\mu\nu\sigma\lambda}
R^{\mu\nu\sigma\lambda}$ this yields ${\cal D}^{-1}R
\approx-\frac{1}{\varepsilon}-\frac{\gamma}{\varepsilon^2M_p^2}
\left(1+\frac{1-\eta}{\varepsilon\eta(2\eta-1)}\right)^{-1}
R_{\mu\nu\sigma\lambda}R^{\mu\nu\sigma\lambda}/R$
(if $|\varepsilon|>\frac{1-\eta}{\eta(2\eta-1)}$),
whereas in the opposite limit $R/R_{\mu\nu\sigma
\lambda}R^{\mu\nu\sigma\lambda}\to 0$ one finds ${\cal D}
^{-1}R\to 0$.} The solution (\ref{25}), (\ref{26}) is an
attractor in the space of general solutions of eqs.
(\ref{13}-\ref{21})
 for $t\to \infty$. For a smooth transition
between the radiation and matter-dominated epoch the
particular solution with the asymptomatic behaviour
(\ref{16b}), (\ref{15}) should be selected such that it
connects continuously with the value of $\varphi=0$
before the transition. Comparing eq. (\ref{26}) with the critical
density in Einstein gravity $\rho_c^{(E)}$, we find that the
value of $\Omega$ for a zero curvature $(k=0)$ universe
as predicted by inflation is
\be\label{39}
\Omega=\frac{\rho_c}{\rho_c^{(E)}}=\frac{2\tau^2-1}{3\tau^2-1}(1-\frac
{\tau^2}{2\epsilon})\ee
For positive $\epsilon>\frac{1}{2}\tau^2$ one finds $0<\Omega<
\frac{2}{3}$ and also for negative $\epsilon<\frac{1}{2}-
\tau^2$ the effective critical density turns out to be smaller
than one. The value preferred by present observation $\Omega\approx
0.3-0.5$ seems well compatible with nonlocal gravity.

In conclusion, the only discrepancy with observation
for the simple model of nonlocal gravity (6) appears to
be nucleosynthesis. The extent
to which the effective Planck mass differs between nuclosynthesis
and today depends, however, crucially on the precise form
of the regularized operator ${\cal D}$: Due to the dilatation
anomaly (running couplings) and presence of mass corrections
in the equation of state the curvture scalar does
actually not vanish exactly during the radiation-dominated
epoch.  Therefore the effective Newton's constant could
also deviate from $1/M^2_p$ for this
period. Furthermore, generalized nonlocal terms of the type
\be\label{40}
{\cal L}_{nl}=\sqrt g R^{\mu\nu}({\tilde{\cal D}}^{-1}
)_{\mu\nu\sigma\lambda}R^{\sigma\lambda}\ee
with $\tilde{\cal D}$ an appropriately regularized differential
operator could also cure the unboundedness
of the euclidean effective action. (Eq. (6) corresponds
to a special case of $\tilde{\cal D}$.) Again, there will be
no changes in the gravitational solutions for empty
space since $R_{\mu\nu}=0$. On the other hand $R_{\mu\nu}$
differs from zero during the radiation-dominated
period in early cosmology, thus leading to an effective value of the
Planck mass different from $M_p$ during nucleosynthesis.
Without a detailed investigation of possible
forms of $\tilde{\cal D}$ in eq. (\ref{40}) it seems difficult
to exclude nonlocal gravity on the basis of nucleosynthesis.
For the moment there seem to be too many free parameters
such that the effective values of the Planck mass
during the radiation and matter dominated periods remain
essentially undetermined. Nucleosynthesis can then be
interpreted as a ``measurement'' of the difference in
$M_p^{eff}$ between the radiation-dominated cosmology
and the Newtonean value (18). Within certain assumptions
a determination of $\Omega$ can play a similar role for the
``measurement'' of $M_p^{eff}$ during matter dominated
cosmology.
It is striking that it seems not possible
to establish the validity of the Einstein
equations for long-distance gravity using
only the principle of general covariance
and phenomenological considerations based on
the presently available tests of
general relativity. On the other hand,
some models of nonlocal gravity, especially involving
generalizations of ${\cal L}_{nl}$ beyond (\ref{40}),
may lead to more drastic observable modifications
of phenomenology.

Of course, the most
crucial question remains if quantum gravity really leads
to an effective action of the type (\ref{40}). A one-loop
computation in pure Einstein gravity gives no indication
in this direction \cite{LOOP} since the nonlocalities appear
here only logarithmic.
The compatibility  of our simple
model of nonlocal gravity with observation should motivate further
non-perturbative studies of long-distance quantum gravity,
perhaps based on exact flow equations \cite{Rev} and not
assuming necessarily a ``classical action'' of the pure
Einstein type. We hope that this work motivates observational and
experimental efforts to look for a possible curvature dependence
of the effective Newton's constant.

\end{document}